\documentstyle[osa]{revtex}

\begin{document}
\title{On the first-order phase transitions in a bistable large-spin systems ($%
Mn_{12}Ac$)}
\author{V V Makhro\thanks{%
E-mail: maxpo@mailexcite.com}}
\address{Department of Physics, Bratsk Industrial Institute, Bratsk, 665709 Russia}
\maketitle

\begin{abstract}
We have shown that recent report\cite{first} concerning the first-order
phase transitions in the systems described by the Hamiltonian ${\cal H}%
=-DS_z^2-H_xS_x$ is inaccurate. A kinetic numerical method for making
calculations of the transition rate in a bistable system as a function of
temperature has been evolved and a new critical value of the transition
field ($h_x=0.891$) has been found. A physical mechanism caused transition
from quantum to classical behavior has been discussed.

\end{abstract}

The thermally behavior of the escape rate is the main indicator for the
search of a quantum phenomena in the large-spin mesoscopic systems. The
system that attracted the most recent attention is the $Mn_{12}$ acetate ($%
Mn_{12}O_{12}(CH_3COO)_{16}(H_2O)_4)\cdot 2CH_3COOH\cdot 4H_2O$). This
crystal has been synthesized by {\it Lis}\cite{lis} and its magnetic
properties was investigated at first by {\it Sessoli et al}\cite{ses}. The
molecules of $Mn_{12}Ac$ effectively behave as magnetic clusters of spin $%
S=10$ and are characterized by very strong uniaxial anisotropy ${\cal H}%
=-DS_z^2$ with $D\simeq 0.75k_B$. The first observations of the tunneling of
a magnetization in $Mn_{12}Ac$ were carried out by {\it Novak and Sessoli}%
\cite{novak}. They found that the relaxation rate of $Mn_{12}Ac$ followed
the Arrhenius law, $\Gamma \sim \exp (-U_{eff}/T)$, where $U_{eff}$ is
height of the energy barrier between the two metastable states $\pm S$, but
for the certain values of the external longitudinal field the peaks raised.
Such a peaks were interpreted as the result of the resonant thermally
assisted tunneling between the levels near the top of the barrier\cite{novak}%
. Later, this phenomena attracted great experimental and theoretical
interest. The successful dynamical theory of thermally activated tunneling
in $Mn_{12}Ac$ was proposed by {\it Garanin and Chudnovsky}\cite{second}.

However, the crossover from thermal to quantum regime for the escape rate
has not yet been satisfactorily explained. Thermal activation and tunneling
are usually considered as competing processes; as a result, one might be lad
to think that discussed mechanisms works in different temperature regions:
tunneling would be observable only at extremely low temperatures, whereas at
higher temperatures tunneling would be suppressed by the thermal activation.
Using this approach one can divide the temperature dependence of the escape
rate onto two regions - high-temperature region, where escape rate follows
the Arrhenius law, $\Gamma \sim \exp (-U_{eff}/T)$ , and low-temperature
region, where the transitions is purely quantum, $\Gamma \sim \exp (-B),$
with $B$ independent on $T$. Formally, one can to equate the right-hand
parts of this equations and found in such manner transition temperature $%
T_c\sim U_{eff}/B$ \cite{first}. In some sense, such a temperature may be
called as a temperature of the{\it \ ''}phase transition'' from classical%
{\it \ }to quantum behavior{\it . }Whereas the two regimes smoothly join,
one can call this situation the{\it \ }second-order phase transition{\it \ }%
\cite{larkin}. {\it Chudnovsky and Garanin}\cite{first} recently made an
attempt to describe quantum-classical transitions in large-spin system more
detailed. They consider a spin Hamiltonian

\begin{equation}
{\cal H}=-DS_z^2-H_xS_x
\end{equation}
with $S>>1$. Such a model is a good approximation for the molecular magnets $%
Mn_{12}Ac$. The states $\pm S$ are separated by the energy barrier $U=DS_z^2$%
. Overcoming this barrier is possible in the following ways: by absorption
of an external field's energy, or due to thermal activation, or via
tunneling. Further, authors of Ref.\cite{first} reduced the initial spin
problem to the particle problem using the mapping method\cite{zas}. The
equivalent particle Hamiltonian will be 
\begin{equation}
{\cal H}=-\frac{\nabla ^2}{2m}+U(x)
\end{equation}
with

\begin{equation}
U(x)=\left( S+1/2\right) ^2D(h_x^2\sinh ^2x-2h_x\cosh x)\text{,}
\end{equation}
and $m\equiv \frac 1{2D}$ , $h_x\equiv \frac{H_x}{(2S+1)}$ .

The function $U(x)$ in (3) described typical double-well potential (see Fig.
1) for all $h_x:0<h_x<1$. Only in boundary points $h_x=1$ and $h_x=0$ this
potential change the form. At $h_x=1$ the barrier completely disappeared,
and when $h_x\rightarrow 0$ the width of barrier tends to infinity.

But authors of \cite{first} used for the further analysis instead of $U(x)$
its expansion with fourth-order accuracy 
\begin{eqnarray}
V(x) &=&-2\left( S+\frac 12\right) ^2Dh_x+\left( S+\frac 12\right) ^2D\left(
h_x^2-h_x\right) x^2+  \nonumber \\
&&\left( S+\frac 12\right) ^2D\left( \frac 13h_x^2-\frac 1{12}h_x\right)
x^4+O(x^6)\text{.}
\end{eqnarray}
For such a potential, with lowering of the transverse field below the
critical value of $h_x=1/4$ the fourth-order term change the sign and became
negative, and the potential ''overturned''. The barrier between opposite
metastable states completely disappeared in this case, and what is more,
metastable states disappeared too (see Fig. 2). It is difficult to believe
that such situation corresponds to any physics. In reality, it means that
approximation $U(x)$ by $V(x)$ is invalid. Authors of Ref.\cite{first}
alleged that field $h_x=1/4$ corresponds to the first-order phase transition
between quantum and classical types of the escape (see also Ref.\cite{thr}).
But it is easy to see, that truncation of the high-order terms in (4) leads
to losing of the physical information. With an increasing of the accuracy of
the expansion, the magnitude of a ''critical field'' rapidly decreased: if
one take into account six-order term, then critical value will be $h_x=1/24$
etc. In the limit we have the exact potential (3) and the critical field $%
h_x=0$.

Thus, it is clear that potential (4) is inadequate to the discussed problem
and complete analysis of the problem for the initial potential (3) is
necessary. For the solving of the described problem we propose the next
numerical scheme.

Let us consider a particle in potential (3). We assume that particle is in
thermal equilibrium with the bath. We consider an ensemble containing $N$
such particles. When the energy of the particle $E$ less then $U_0$ it will
be localized in the well into one of the stationary states $E_n$, otherwise,
it will be unbound. Physically, first case corresponds to ''pinned''
magnetization, and, at last case the escape occurred and magnetization
oscillated between ${\bf z}$ and $-{\bf z}$ directions with a frequency 
\[
\varpi _{ext}=\frac \pi {\sqrt{\frac m2}\int_{x_1(E)}^{x_2(E)}\frac{dx}{%
\sqrt{E-U(x)}}}\text{,} 
\]
where $x_{1,2}(E)$ are turning points for the particle oscillating inside
the potential $U(x).$ For the values $E<U_0$ the escape can took place only
via tunneling. We will try to determine the fraction of the particles from
the ensemble $N$ which overcoming the barrier for the given temperature $T$
due to both the thermal activation and the tunneling. The number of
particles overcoming the barrier due the thermal activation will be 
\[
N_{therm}=N\int_{U_0}^\infty \frac 2{\sqrt{\pi }}(k_BT)^{-1.5}\sqrt{E}\exp (-%
\frac E{k_BT})dE\text{,} 
\]
where $U_0=U_{\max }-U_{\min }$, $U_{\max }=-2S^2Dh_x,$ $U_{\min
}=-S^2D(1+h_x^2)$ . In numerical calculations, for the upper bound we of
course make the substitution $\infty \rightarrow U_r$ where $U_r$ was
adjusted so as to lead to the neglect of number of particles outside the
interval for each given value of the temperature. Residuary number of
particles $N_{well}$ will distribute among the stationary levels in
according with 
\begin{equation}
n_i=N_{well}\frac{\exp (-E_i/k_BT)}Z
\end{equation}
where $Z=\sum\limits_i\exp (-E_i/k_BT)$ is the partition function. For each $%
E_{i\text{ }}$we calculate the probability of the tunneling using the WKB
approximation for the lowest levels 
\begin{equation}
\Omega _i=\frac 1\pi \exp (-\frac 1{2D}\int_{-A_i}^{A_i}((S+\frac 12%
)^2D(h_x^2\sinh ^2x-2h_x\cosh x)-E_i)dx,
\end{equation}
and for the levels near the top of barrier we use the approximation of $U(x)$
by the parabolic potential; in this case probability will be\cite{kem} 
\begin{equation}
\Omega _i=\frac 1{1+\exp (-2\pi E_i/\sqrt{2D})}\text{.}
\end{equation}
$A_i$ we found numerically from the equation $U(x)=E_i$. The number of
particles overcoming the barrier via tunneling will be $N_{tun}=\sum%
\limits_in_i\Omega _i$ and $N_{tot}=N_{tun}+N_{therm}$ will be total number
of the particles which overcome the barrier, and in conclusion, the
effective barrier transparency $P$ one can to determine as $P=N_{tot}/N.$

For the calculation of the spectrum of the particle we use approximation $%
U(x)$ by well-known potential 
\begin{equation}
W(x)=C(e^{-2\alpha x}-2e^{-\alpha x})
\end{equation}
and assume the barrier width for the $U(x)$ tends to infinity (see Fig.3).
In such unperturbed case the spectrum of the particle inside one of the
wells will be\cite{lan} 
\begin{equation}
E_i=-A\left( 1-\alpha \sqrt{D/A}(i+\frac 12)\right) ^2,
\end{equation}
where prefactor $A\approx 1.1U_0$ and $\alpha $ changes from $1.3$ for $%
h_x=0.1$ to $2.4$ for $h_x=0.8$. In real potential $U(x)$ another well of
course will perturb the spectrum. But the level's splitting will be
essential only for levels near the top of barrier, which has very weak
population in discussed temperature region (up to $1K$). Thus, for the
preliminary estimation this approximation seems to be admissible. Let us
note that in any case there exist minimal depth of the well $U_{0\min }$. If 
$U_0<U_{0\min }$ discrete spectrum inside the well will be lacking. For the
potential (8), one can found $U_{0\min }$ from the condition\cite{lan} $%
A=D/4 $, that corresponds to critical value of the transverse field $%
h_{xcr}=0.891. $ It is clear, that for $h_x\geq h_{xcr}$ the escape may
occur only due to thermal activation. Therein, one can say about transition
from quantum to classical behavior. Such a transition will be abrupt and may
be called ''first-order'' phase transition, but any analogies with the
Landau theory of phase transitions not stand up to criticism.

In conclusion, let us discussed some results of the numerical calculations.
From the data in Table 1 one can see that the spectrum of the particle in
the potential (3) is characterized by the small number of levels. In
discussed temperature region for small $h_x$ all the particles condensed, as
a rule, at the ground-state level, where the tunneling probability is
negligible. The population of the higher levels tends to zero, so, the
effective transparence of the barrier tends in this case to zero too. With
increasing of $h_x$ the population of higher levels where tunneling
transparence is considerable, increased too, and effective transparence
amount to appreciable values. For the values $h_x\geq 0.891$ the overcoming
the barrier can be possible only due to thermal activation. In Fig. 4 and 5
we plotted dependences of the effective barrier transparence on T for
different values of the transverse field $h_x.$ The difference in the
behavior of the curves for $h_x<h_{xcr}$ and $h_x\geq h_{xcr}$ is evident.

Table 1.

Some characteristics of the escape process at $T=1K$ for the different
values of the transversal field $h_x$ ($N=100$).

\begin{tabular}{lllllll}
$h_x$ & $0.1$ & $0.3$ & 0.5 & 0.6 & 0.7 & 0.891 \\ 
Number of levels & 7 & 5 & 3 & 2 & 1 & 0 \\ 
Population of the lowest level & 100 & 100 & 99.993 & 99.85 & 99.61 & - \\ 
Population of the highest level & 0 & $1.615\cdot 10^{-11}$ & $1.871\cdot
10^{-4}$ & 0.147 & 99.61 & - \\ 
Tunneling probability for the lowest level & 0 & 0 & $1.344\cdot 10^{-5}$ & $%
3.426\cdot 10^{-3}$ & 0.334 & - \\ 
Tunneling probability for the highest level & 0.261 & 0.318 & 0.325 & 0.329
& 0.334 & - \\ 
Probability of the thermal activation & 2.23$\cdot 10^{-20}$ & 3.15$\cdot
10^{-11}$ & 3.65$\cdot 10^{-5}$ & 2.80$\cdot 10^{-3}$ & 0.39 & 0.62
\end{tabular}

Figure captions

Fig. 1. Reduced effective potential $U(x)$ for the spin system $Mn_{12}Ac$. $%
h_x=1$ (1), 0.75 (2), 0.5 (3), 0.25 (4), 0.15 (5).

Fig. 2. ''Truncated'' potential $V(x)$. $h_x=$ 1 (1), 0.75 (2), 0.5 (3),
0.25 (4), 0.15 (5).

Fig. 3. Potential $U(x)$ and approximating potential $W(x)$ for $h_x=0.1.$

Fig. 4. Dependence of the effective barrier penetrability on $T$ for $%
h_x=0.7 $ (mixed thermal activated and tunneling escape).

Fig. 5. Dependence of the effective barrier penetrability on $T$ for $%
h_x=0.891$ (pure thermal activated escape).

\end{document}